%% file: bare_jrnl_new_sample4.tex
\documentclass[lettersize,journal]{IEEEtran}
\usepackage{amsmath,amsfonts}
\usepackage{algorithmic}
\usepackage{algorithm}
\usepackage{array}
\usepackage[caption=false,font=footnotesize,labelfont=rm,textfont=rm]{subfig}
\usepackage{textcomp}
\usepackage{stfloats}
\usepackage{url}
\usepackage{verbatim}
\usepackage{graphicx}
\usepackage{cite}
\usepackage[table,xcdraw]{xcolor}
\usepackage{booktabs} 
\begin{document}

\title{\textcolor{black}{New Characteristics and Modeling of 6G Channels: Toward a Unified Channel Model for Standardization}}

\author{Huiwen Gong, Jianhua Zhang, Yuxiang Zhang, Guangyi Liu
        % <-this % stops a space
\thanks{Huiwen Gong, Jianhua Zhang, Yuxiang Zhang are with the State Key Laboratory of Networking and Switching Technology, Beijing University of Posts and Telecommunications, Beijing 100876, China (email: birdsplan@bupt.edu.cn; jhzhang@bupt.edu.cn; zhangyx@bupt.edu.cn;)
Guangyi Liu is with the Future Research Laboratory, China Mobile Research Institute, Beijing 100053, China (e-mail: liuguangyi@chinamobile.com;).}% <-this % stops a space
\thanks{Manuscript received xxx, xxx; revised xxx, xxx.}}

\markboth{Journal of \LaTeX\ Class Files,~Vol.~14, No.~8, August~2021}%
{Shell \MakeLowercase{\textit{et al.}}: A Sample Article Using IEEEtran.cls for IEEE Journals}

\IEEEpubid{0000--0000/00\$00.00~\copyright~2021 IEEE}

\maketitle

\begin{abstract}
As 6G research advances, the growing demand leads to the emergence of novel technologies such as \textcolor{black}{integrated sensing and communication (ISAC), new antenna arrays like extremely large MIMO (XL-MIMO) and reconfigurable intelligent surfaces (RIS), along with multi-frequency bands (new mid-band, above 100 GHz).} Standardized unified channel models are crucial for research and performance evaluation across generations of mobile communication, but the existing 5G 3GPP channel model based on geometry-based stochastic model (GBSM) requires further extension to accommodate these 6G technologies. 
\textcolor{black}{In response to this need, this article first investigates several distinctive channel characteristics introduced by 6G technologies,}
such as ISAC target radar cross-section (RCS), sparsity in the new mid-band, and others. Subsequently, an extended GBSM (E-GBSM) is proposed, integrating these characteristics into a unified modeling framework. The proposed model not only accommodates 6G technologies with flexibility but also maintains backward compatibility with 5G, ensuring a smooth evolution between generations. Finally, the implementation process of the proposed model is detailed, \textcolor{black}{and experiments and simulations are presented to validate its effectiveness and accuracy}, providing support for 6G channel modeling standardization efforts.
\end{abstract}


\section{Introduction}
The sixth generation (6G) of wireless communication technology aspires to redefine connectivity, offering transformative advancements such as higher data rates, ultra-low latency, and ubiquitous coverage \cite{society_6G}. Given the complexity of 6G systems, standardization serves as a cornerstone for integrating diverse technological advancements. In particular, global standards play a pivotal role in ensuring compatibility, interoperability, and scalability across systems and regions. Notably, since 2020, the International Telecommunication Union (ITU) has spearheaded 6G research, targeting deployment by 2030, and finalized its 6G Development Framework in 2023 \cite{ITU-R_M2160}. Concurrently, the World Radiocommunication Conference (WRC) is addressing spectrum requirements, with allocations expected by 2027. The 3rd Generation Partnership Project (3GPP) has also launched Release 19 to initiate 6G standardization efforts.
As with previous generations, channel modeling remains foundational, providing the models essential for uniform benchmarking, system design, and performance evaluation.

The evolution of channel models from 1G to 6G is summarized in Table \ref{table1}, highlighting the transition from empirical models to standardized approaches. For 2G and earlier generations, empirical path loss models such as the Okumura-Hata model and the COST-231 model were predominant. Starting from 3G, standardized channel models have been introduced for each generation of mobile communication. The 3G era saw the introduction of the \textcolor{black}{spatial channel model (SCM)}, combining statistical modeling with spatial correlation to support frequency-selective fading. With the widespread adoption of MIMO technology, 4G introduced the \textcolor{black}{spatial channel model extension (SCME)} and WINNER models, which further enhanced spatial modeling capabilities. For 5G, the \textcolor{black}{geometry-based stochastic model (GBSM)} integrated geometric and stochastic properties\cite{38901}, while the 3D-GBSM extended channel representation into the vertical domain.

\input{table1}

6G aims to transform communication systems by expanding application scenarios and adopting advanced technologies to meet unprecedented performance demands. The IMT-2030 Framework \cite{ITU-R_M2160} outlines six 6G use cases, building on three primary 5G scenarios, with 15 performance indicators surpassing 5G in sensing, positioning, data rates, and coverage. Several novel technologies have been proposed to achieve these goals. \textcolor{black}{Integrated sensing and communication (ISAC)} combines sensing and communication functions for precise positioning and autonomous driving. Advanced antenna technologies, such as \textcolor{black}{extremely large MIMO (XL-MIMO)} and \textcolor{black}{reconfigurable intelligent surfaces (RIS)}, enhance spectral efficiency, capacity, and reliability. XL-MIMO utilizes extensive antenna arrays, while RIS actively manipulates electromagnetic waves to improve coverage. To support these capabilities, new frequency ranges, such as the new mid-band, provide the bandwidth and spectrum resources required to achieve 6G’s peak data rates \cite{6G_wide_band_shafi}.

\IEEEpubidadjcol

While the 5G channel model, grounded in GBSM, has successfully supported key technologies like MIMO and millimeter-wave communication, it now cannot be directly applied to 6G due to expanded technical requirements, broader frequency ranges, and more diverse scenarios introduced by 6G technologies. ISAC systems, for instance, necessitate target modeling to capture the characteristics of both communication and sensing channels\cite{ISAC_mag_wang}. XL-MIMO’s large antenna arrays introduce spatial non-stationarity (SnS) and near-field effects\cite{XLMIMO_yuan}, significantly altering channel behavior. RIS-assisted communications present challenges in modeling the \textcolor{black}{transmitter} (Tx)-RIS and RIS-\textcolor{black}{receiver} (Rx) channel coupling, while new frequency bands, such as mid-band \cite{Sparsity_miao}, millimeter-wave, and \textcolor{black}{sub-terahertz (sub-THz)}, exhibit sparse multipath propagation\cite{Sparsity_ximan}. 

These diverse technologies introduce unique propagation characteristics, complicating the standardization process if individual models are developed for each. Such fragmentation could hinder the establishment of unified standards, leading to inconsistent evaluations and ultimately slowing down standardization efforts. To address these challenges, Release 19 discussions advocate for a unified model \cite{7-24_3GPP_M}, as reflected in the consensus to \textit{design a unified model to explicitly reflect the new properties of near- and existing properties of far-field.} A unified framework ensures consistency across diverse deployment scenarios, streamlining system design, spectrum allocation, and performance evaluation, thereby accelerating 6G advancements. Moreover, the GBSM framework itself offers notable scalability and adaptability. Despite its current limitations in capturing new scenarios and technologies, its long-standing application from 4G to 5G has demonstrated robust accuracy and reliability. This flexibility positions GBSM as a strong foundation for extension, capable of supporting the diverse demands of 6G, rather than requiring an entirely new modeling paradigm.

In response to these challenges, this article investigates the channel characteristics introduced by key 6G technologies and, drawing on discussions from the 3GPP Technical Specification Group Radio Access Network Working Group 1 (TSG RAN1), \textcolor{black}{proposes a channel modeling framework based on an \textcolor{black}{extended GBSM (E-GBSM)}.
This framework moves toward a unified modeling approach by integrating the distinctive features of ISAC, advanced antenna technologies, and new frequency bands (new mid-band and above 100 GHz), while ensuring compatibility with existing standards.}
By extending current models, it maintains backward compatibility, provides a clear implementation roadmap, and addresses critical requirements such as wideband support and multi-technology evaluation.

\section{\textcolor{black}{New channel characteristics for 6G technologies}}
\begin{figure*}
    \centering
    \includegraphics[width=\linewidth]{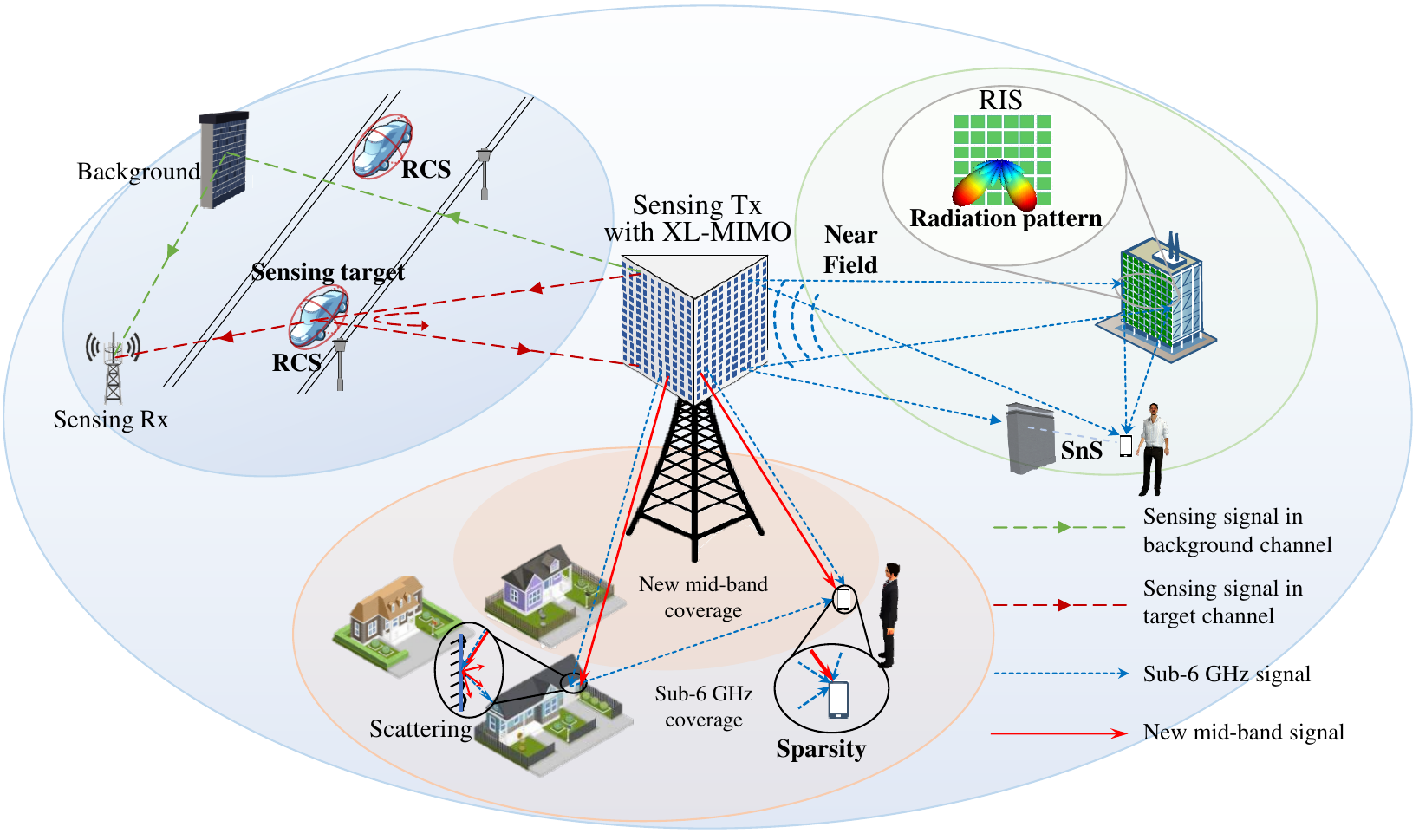}
    \caption{Channel characteristics illustration for new technologies in 6G.}
    \label{fig:illustration}
\end{figure*}

Developing effective 6G channel models requires capturing the unique propagation behaviors introduced by diverse enabling technologies. Understanding the channel characteristics of these technologies is a critical first step, as it provides the foundation for designing accurate and adaptable channel models. Based on channel measurement results, this section outlines the key characteristics of 6G novel communication technologies: ISAC, new antenna technologies, and new frequency communication. These findings serve as the basis for the subsequent modeling methods. Fig. \ref{fig:illustration} illustrates the channel characteristics considered.

\subsection{ISAC}
Compared to traditional communication channels, ISAC channels emphasize the behavior and characteristics of sensing targets. Measurement studies focus on three aspects: (1) \textcolor{black}{radar cross-section (RCS)} properties of sensing targets, (2) modeling of the target channel and its sub-channels (Tx-target and target-Rx), and (3) correlations between sensing target channels and communication channels.

The first aspect involves the investigation of RCS properties. Although extensively studied in radar systems, ISAC targets differ in frequency bands and scenarios, necessitating further research \cite{ISAC_liufan}. Typical targets include \textcolor{black}{unmanned aerial vehicle (UAV)}, human, and vehicle, relevant to applications such as UAV detection and autonomous driving assistance. Measurements reveal distinct RCS distributions, e.g., \textcolor{black}{small UAVs and humans exhibit weak angular dependence, while vehicles show four peaks corresponding to their sides.} RCS also varies with frequency and distance, stabilizing at longer ranges. Certain studies highlight potential correlations between bistatic and monostatic RCS within specific angular ranges, simplifying modeling.

For target channels, standardized models typically treat them as concatenated sub-channels. On a large scale, path loss is calculated by summing the sub-channel losses (in dB), accounting for RCS and omnidirectional antenna effects. On a smaller scale, the convolution-based model characterizes multipath propagation, though specific methodologies remain under discussion. Fixed environmental objects (e.g., walls) further influence signal behavior within the target channel.

The third focus examines shared scatterers, which introduce correlations between sensing and communication channels. In ISAC systems, shared scatterers, interacting with signals in both channels, exhibit similar propagation characteristics (e.g., angles and delays), posing challenges for integrated channel modeling.

\subsection{New antenna}

The channel characteristics of advanced antenna technologies, such as XL-MIMO and RIS, focus on near-field effects, SnS, and multi-segment concatenated channels in RIS systems.

For XL-MIMO, near-field spherical waves and SnS present unique modeling challenges. Unlike traditional MIMO, where far-field plane wave assumptions hold, XL-MIMO systems often place users in the near-field due to the larger array aperture, where Rayleigh distance increases quadratically with aperture size. Measurements reveal that near-field signals deviate from plane wave assumptions, with nonlinear phase variations and amplitude differences across antenna elements\cite{XLMIMO_yuan}. While this complicates beamforming, the additional spatial degrees of freedom can enhance channel capacity.

SnS arises from XL-MIMO's larger physical aperture, where elements experience distinct multipath conditions. Measurements show variations in multipath visibility and power, driven by blockage and transitions between direct and reflected or diffracted paths, further complicating channel modeling. 

RIS typically operates as a passive relay between the Tx and Rx, forming a multi-segment concatenated channel (Tx-RIS-Rx). The relationship between the concatenated channel and its two sub-channels is determined by RIS parameters such as array size, control methods, and phase-shift resolution.

\textcolor{black}{Similar to the Tx-target and target-Rx sub-channels in ISAC systems, RIS channel modeling also follows a concatenation approach. Large-scale path loss initially assumed to scale with the sum of squared sub-channel distances was later corrected to scale with their product, supported by experimental measurements. On a small scale, RIS channel models typically treat the concatenated channel as a convolution of the two sub-channels.} For example, if the Tx-RIS channel has M paths and the RIS-Rx channel has N paths, the concatenated channel will have $M \times N$ paths, with delays and powers determined by the sum and product of the respective sub-channel characteristics.

The control of signals by RIS can be characterized using an equivalent radiation pattern. RIS models typically fall into two categories: (1) models based on mathematical fitting of the radiation pattern, and (2) electromagnetic-based models. The former uses simplified antenna theory expressions, such as the $cos^q \theta$ form, to fit the RIS element's radiation pattern, while the latter employs methods such as physical optics or RCS approximations to compute the RIS radiation gain.

\subsection{New frequency}
This article focuses on two new frequency bands: the new mid-band, which was approved for study in 3GPP Release 19 in 2023, and the frequencies above 100 GHz, which the WRC has highlighted for potential terrestrial communication applications. Both bands are higher than existing commercial frequencies and share similar characteristics, such as channel sparsity. Additionally, these new bands may be applied in emerging scenarios like sub-urban macro (SMa) environments, where channel parameters are expected to exhibit features distinct from current models.

\textcolor{black}{As carrier frequency increases and the signal wavelength decreases, two propagation effects emerge. First, the diffraction capability of shorter wavelengths diminishes, resulting in a stronger reliance on line-of-sight (LOS) and a limited number of reflected paths. This reduces the number of resolvable clusters in the environment. Second, the smaller wavelength enlarges the relative scale of surface irregularities on objects, leading to more angularly dispersed and lower-power scattering components. As a result, the intra-cluster power tends to concentrate in a few dominant reflected paths.
Despite these frequency-dependent characteristics, current 3GPP channel models do not reflect such sparsity. Across a wide frequency range—from 0.5 to 100 GHz—the models assume a fixed number of clusters for a given scenario and assign equal power to 20 paths within each cluster.}
Various metrics, such as channel rank or spatial degrees of freedom, could quantify sparsity. This article adopts the Gini Index, as proposed in \cite{Sparsity_ximan}, to measure sparsity, where a Gini Index closer to 1 (within the 0-1 range) indicates greater sparsity. Channel measurements from \cite{Sparsity_ximan} across centimeter-wave, new mid-band, and THz bands show that the Gini Index increases with frequency, confirming that higher frequencies exhibit more pronounced sparsity. 

\begin{figure*}[t]
    \centering
    \includegraphics[width=\linewidth]{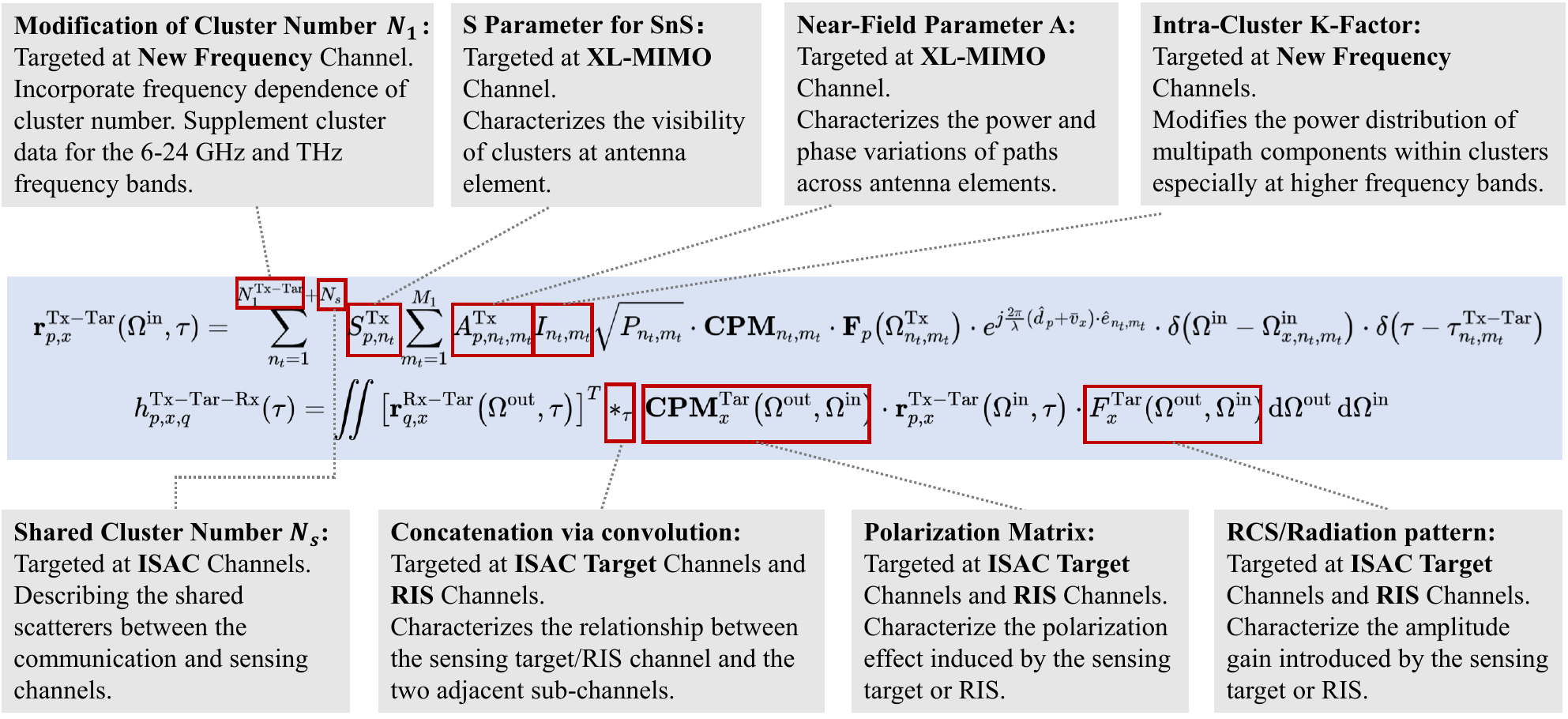}
    \caption{\textcolor{black}{The illustration of the E-GBSM for 6G.}}
    \label{fig:feture}
\end{figure*}

\section{\textcolor{black}{Channel modeling for 6G}}
\subsection{Consideration for 6G channel model}
The unique requirements introduced by 6G technologies for channel modeling can be summarized as follows:

\textit{Support for evaluating multiple technologies.} While the 5G channel model accommodates several features like spatial consistency, temporal evolution, and blockage effects (e.g., \textcolor{black}{vehicle-to-everything (V2X)} networks and oxygen absorption for 53-67 GHz), it lacks provisions for many 6G technologies. For instance, ISAC requires models that capture sensing channels from the Tx to the target and then to the Rx, which are unsupported in current models like 3GPP TR 38.901 \cite{38901}. Similarly, RIS models must represent concatenated Tx-RIS-Rx channels, and XL-MIMO demands adjustments to incorporate non-stationary and near-field effects, which traditional models based on spatial stationarity and far-field assumptions fail to address.

\textit{Support for a wide frequency range.} Achieving 6G’s peak data rates requires both higher frequencies (e.g., THz bands \cite{THz_guan}) and better utilization of underused bands (e.g., new mid-band). \textcolor{black}{Current models extrapolate new mid-band parameters from FR1 and FR2 bands \cite{mid_band_rappaport} and lack representation of the frequencies above 100 GHz.} Moreover, they do not account for frequency-dependent small-scale fading characteristics, such as the reduction in cluster and path counts at higher frequencies.

\textit{Compatibility with existing standards and minimal modifications.} Channel models have evolved by building upon previous generations, integrating new features while maintaining continuity. A 6G model should follow this progression, ensuring continuity in technical evaluations and enabling reuse of established knowledge, tools, and methodologies. The GBSM framework’s scalability allows for the inclusion of new channel features with minimal changes, facilitating consistent modeling for new technologies.

\textit{Accuracy based on empirical data.} Parameters in the standardization channel model are statistically derived from extensive measurements across diverse scenarios, such as \textcolor{black}{urban micro (UMi), urban macro (UMa), rural macro (RMa), and indoor hotspot (InH). To address expanded use cases, 5G added scenarios like indoor factory (InF).} Similarly, 6G introduces novel environments that necessitate additional measurement campaigns to ensure accurate parameterization, thereby guaranteeing model reliability and relevance.

\subsection{A unified channel modeling framework toward standardization}
This section presents a unified channel modeling framework tailored for 3GPP standardization, developed under the principles outlined in the preceding discussion. The proposed framework builds upon the E-GBSM, ensuring seamless compatibility with existing standardized models while necessitating only minimal adjustments. Designed to address the distinct requirements of 6G technologies, it facilitates simulations for four key advancements: ISAC, XL-MIMO, RIS, and new frequency communications. Grounded in measurement data, the framework addresses diverse deployment scenarios with relevance and adaptability.

\begin{figure*}
    \centering
    \includegraphics[width=0.95\linewidth]{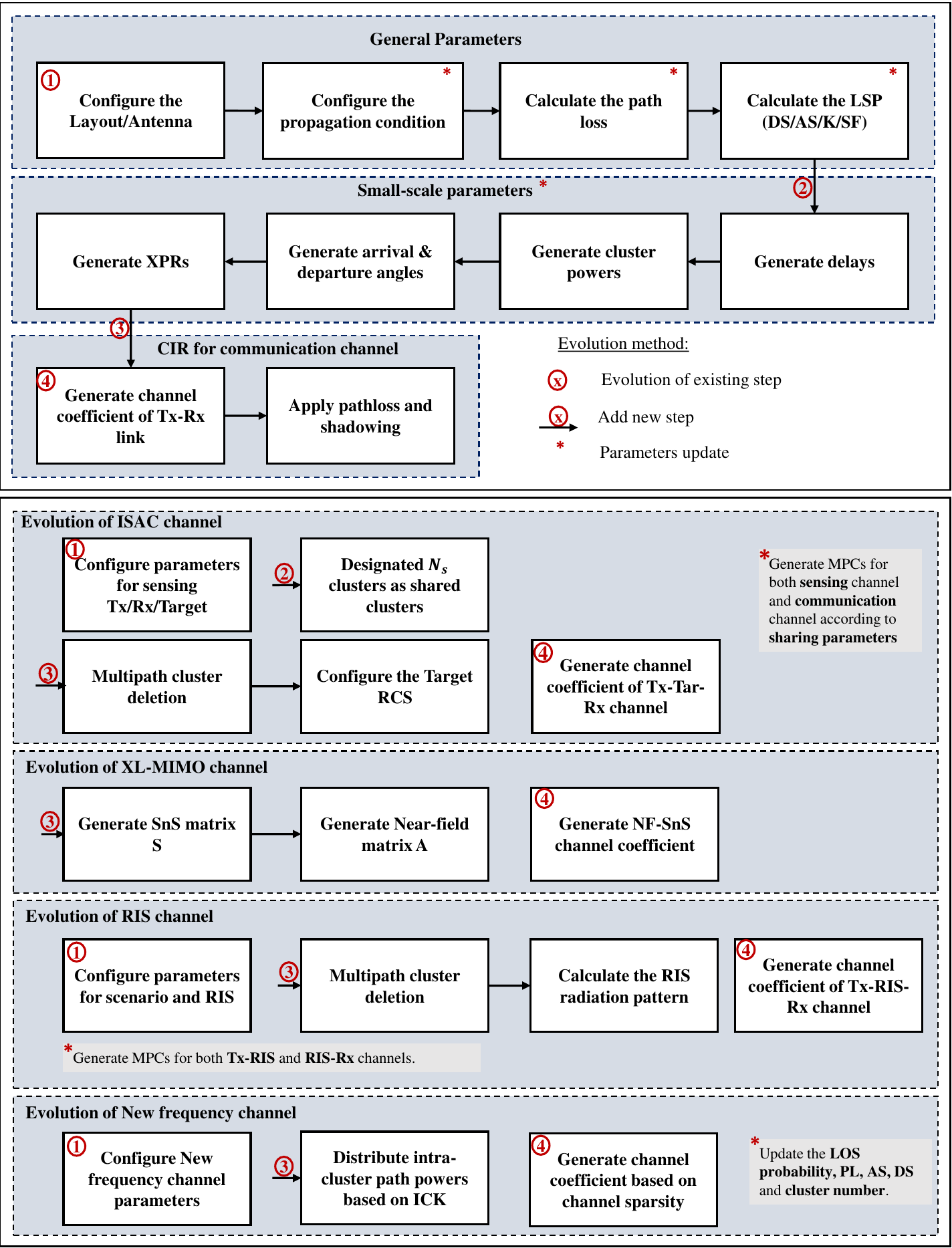}
    \caption{\textcolor{black}{Implementation of the unified channel model based on E-GBSM within the 3GPP framework. The upper figure illustrates the implementation process of the 5G 3GPP model \cite{38901}, while the lower figure depicts the evolution of various channels under the unified channel modeling framework.}}
    \label{fig:step}
\end{figure*}

The unified channel model is illustrated in Fig. \ref{fig:feture}. 
\textcolor{black}{
The first equation characterizes the channel between the $p$th transmit (Tx) antenna and the $x$th intermediate node (either an ISAC target or a RIS panel). Here, $\mathbf{F}_p$ is a $2 \times 1$ vector, and $\mathbf{CPM}_{n_t, m_t}$ is a $2 \times 2$ matrix. The resulting $2 \times 1$ vector $\mathbf{r}_{p,x}^{\mathrm{Tx{-}Tar}}$ represents the angle-domain and delay-domain channel impulse response (CIR) for the two polarization components.
Similarly, for the Target–Rx link, denoted as $\mathbf{r}_{q,x}^{\mathrm{Rx{-}Tar}}(\Omega^{\mathrm{out}}, \tau)$, the expression follows the same structure as in the Tx–Target case, with the following substitutions: Tx $\rightarrow$ Rx, $p \rightarrow q$, $n_t \rightarrow n_r$, and $m_t \rightarrow m_r$.}
The second equation captures the overall Tx-target (or RIS)-Rx channel, reflecting the concatenated channel characteristics influenced by the target or RIS.

The red-boxed terms represent extensions to the 5G GBSM model, accounting for the channel characteristics of the four key 6G technologies. 
First, the frequency dependency of cluster number $N_1$ is addressed, which arises due to 6G's potential support for wideband and new frequency communications. Without an explicit functional form for this dependency, lookup tables are used to update parameters for generating the channel's \textcolor{black}{multipath components (MPCs)}. Unlike the fixed number of clusters in the same scenario as in the 3GPP model, the new model adjusts the cluster count based on measurement results across new mid-band to THz, as detailed in \cite{Sparsity_ximan}.

The shared characteristics in the ISAC channel are captured by the second parameter, $N_s$. In the model, clusters are classified into two types: the first category consists of clusters used exclusively for communication or sensing, with their number denoted as $N_1$; the second category represents shared clusters used for both communication and sensing, corresponding to the shared scatterers. The number of these shared clusters is denoted as $N_s$, and these scatterers maintain certain similarities between the communication and sensing channels during the generation process, such as similar delays, angles, and other MPCs.

Parameter $\mathrm{S}$ $(S^{Tx/Rx}_{p,n})$ characterizes the spatial stationarity of a cluster relative to a specific Tx or Rx antenna element\cite{XLMIMO_yuan}. It introduces the Visibility Region concept from the WINNER model, where a state transition between [0,1] indicates cluster visibility and power variation for a given element. The array is divided into \textcolor{black}{station regions (SRs)}, each with a fixed number of elements and a constant $\mathrm{S}$ value. Measurement data are used to derive the average visibility probability of all clusters for the starting SR, providing the $\mathrm{S}$ parameter. A Markov process then extrapolates $\mathrm{S}$ values for the remaining SRs across the array.

Parameter $\mathrm{A}$ $(A^{Tx/Rx}_{p,n,m})$, is a path-level parameter that captures the phase differences between near-field spherical waves and far-field plane waves across the array \cite{XLMIMO_yuan}. This can be understood as the second-order term in the Taylor expansion of the array steering vector, where far-field assumptions only retain the first-order term. For model implementation, it's essential to differentiate between LOS and non-line-of-sight (NLOS) components. The $\mathrm{A}$ parameter for LOS can be directly computed from the Tx/Rx geometry, while for NLOS components, intermediate scattering objects—particularly the first-bounce scatterer (FBS) and last-bounce scatterer (LBS)—must be taken into account.

The fifth parameter, the \textcolor{black}{intra-cluster K-factor (ICK)}, is frequency-dependent and represents the power ratio between the dominant path within a cluster and other paths. This parameter, ranging between [0,1], is derived from measurements \cite{Sparsity_ximan}. Similar to the Rician K-factor in existing models, the ICK allows for the redistribution of power within a cluster to simulate sparse channels. In practice, some studies model this parameter as a Gaussian-distributed statistic.

The $*$ symbol below Fig. \ref{fig:feture} denotes a convolution operation, indicating that the Tx-target (or RIS)-Rx concatenated channel is the convolution of the two sub-channels in the delay domain. For each sub-channel, the angles on either side of the target/RIS are represented using Dirichlet functions. In the concatenated channel, these Dirichlet functions must be integrated over the RCS (or radiation pattern) of the target/RIS to determine the gain effect on signals incident from and exiting at specific angles.

\textcolor{black}{The parameter $ F^{\mathrm{Tar}} $ represents either the square root of the target's RCS or the equivalent radiation pattern of a RIS. Recent 3GPP discussions on ISAC have converged to a consensus on the specific parametrization of monostatic RCS, which is formulated as:
$F^{\mathrm{Tar}} = A \cdot B_1(\Omega^{\mathrm{out}},\Omega^{\mathrm{in}}) \cdot B_2,$
where $ A $ denotes the total reflected power intensity of the target, $ B_1 $ characterizes the angle-dependent scattering component, and $ B_2 $ represents a statistical distribution term modeling the random fluctuation of RCS.
The term $ \mathrm{CPM}^{\mathrm{Tar}} $ describes the polarimetric interaction between the target and the incident signal. Its mathematical structure aligns with the polarization matrix defined in the 3GPP TR 38.901 standard, adopting a $ 2 \times 2 $ Jones matrix formalism.}

\subsection{\textcolor{black}{Implementation for the unified channel model}}
Fig. \ref{fig:step} shows the link-level simulation implementation of the proposed unified channel model. In particular, the top of Fig. \ref{fig:step} shows the current 5G standard channel model implementation process as defined by 3GPP \cite{38901}, which can generally be divided into three major steps: generation of large-scale parameters such as AS and DS, generation of small-scale parameters like cluster delays, powers, angles, and intra-cluster path angles, followed by the synthesis of the CIR.

\textcolor{black}{The implementation of new model is fully built upon the framework at the bottom of the Fig. \ref{fig:step}}, and achieves the integration of different technologies by updating existing steps, adding new steps, and modifying parameters. The common updates for ISAC, XL-MIMO, RIS, and new frequency communication channels are annotated within each specific implementation step. 

\textit{Evolution of ISAC channel}. In step 1, in addition to configuring the scenario layout and Tx/Rx antenna parameters, further settings are needed for sensing targets, particularly their location, RCS, and velocity.
Unlike conventional communication channels that involve only a Tx-Rx link, ISAC systems also include a Tx-target-Rx sensing link. Thus, MPCs must be generated for three segments: Tx-target, target-Rx, and Tx-Rx. These segments exhibit correlated characteristics, which are addressed in step 2 through the generation of shared scatterers. A portion ($N_s$) of clusters in both the communication and sensing channels are designated as shared, with their parameters adjusted to preserve MPC similarity.
\textcolor{black}{Compared with the authors' prior work, which primarily focused on modeling shared scatterers between communication and sensing channels, the ISAC component in Fig.\ref{fig:step} highlights a distinct feature: the modeling of the sensing channel as the concatenation of the Tx-target and target-Rx channels via RCS.}
Fig.\ref{fig:feture} illustrates the proposed convolution-based concatenation method. However, this approach has a computational complexity of $\mathcal{O}(N^2)$, where $N$ denotes the number of paths in one segment of the channel. To address this, an additional step—step 3—is introduced before computing the CIR in updated step 4. This step prunes excess multipath clusters from the Tx-target and target-Rx segments, thereby reducing the number of clusters and paths involved.

\textit{Evolution of XL-MIMO channel}. In this model, two new steps are added for generating the XL-MIMO channel CIR. The first is the generation of the SnS matrix (parameter S), and the second is the generation of the near-field phase term (parameter A). Detailed methods for generating these two parameters are provided in \cite{XLMIMO_yuan}. It is worth noting that some measurements show that the variation in received power across different array elements due to near-field effects in XL-MIMO is minimal. Therefore, when considering parameter A in the unified channel model, only phase variations are introduced, without incorporating power variations.

\textit{Evolution of RIS channel}. The method for generating RIS-assisted communication channels is quite similar to the target channel in ISAC systems. First, in step 1, the channel scenario parameters are configured alongside RIS-specific parameters, including its position, number of elements, codebook, and normal direction. Similarly, when generating the channel’s MPCs, both the Tx-RIS and RIS-Rx sub-channels need to be considered. However, unlike ISAC, where correlation between the two sub-channels is emphasized, in RIS systems, typically used for coverage enhancement, the MPCs of both sub-channels can be generated independently.
Since the Tx-RIS-Rx concatenated channel CIR is generated using the convolution method described in Fig. \ref{fig:feture}, it inherits the simulation complexity of $\mathrm{O}(N^2)$. This challenge is even more pronounced for RIS, as its equivalent radiation pattern depends on the incident wave, making its computation proportional to the number of MPCs in the sub-channel. To address this, step 3 reduces the number of multipath clusters, significantly mitigating computational overhead. Following this reduction, the RIS radiation pattern is calculated using a physical optics-based approach in the unified model.

\textit{Evolution of new frequency channel}. The modeling of new frequency channels primarily involves updating parameters based on measurement activities. First, \textcolor{black}{the scene-level parameters are configured}. While existing 3GPP standards use lookup tables to generate large-scale parameters for different scenarios, the unified channel model updates these tables for the new mid-band and millimeter-wave frequencies, and adds entries for the sub-THz band based on measurements reported in \cite{Sparsity_ximan}. \textcolor{black}{A significant modification concerns the number of clusters ($N_1$), as measurements consistently report fewer clusters than those in corresponding scenes and frequency bands in the current 3GPP model. This also impacts the generation methods for LOS probability, path loss, and other large and small-scale parameters, as outlined in \cite{Sparsity_ximan}.}
Beyond updating parameter tables, step 3 introduces intra-cluster power redistribution via the ICK, enhancing the strongest path within a cluster to dominate the power and thus introducing sparsity. The ICK coefficient is modeled as a Gaussian-distributed random variable based on measurement fittings, which is then used to allocate intra-cluster power. This method ultimately produces the CIR of the new frequency sparse channel through multipath combination.

\subsection{Simulation of the unified channel model}
Based on the proposed model implementation, we developed a channel simulation platform named BUPTCMCCCMG-IMT2023 \cite{platform}. In this section, we present some simulation results to demonstrate the effectiveness of the unified channel model.

\begin{figure}
    \centering
    \includegraphics[width=0.95\linewidth]{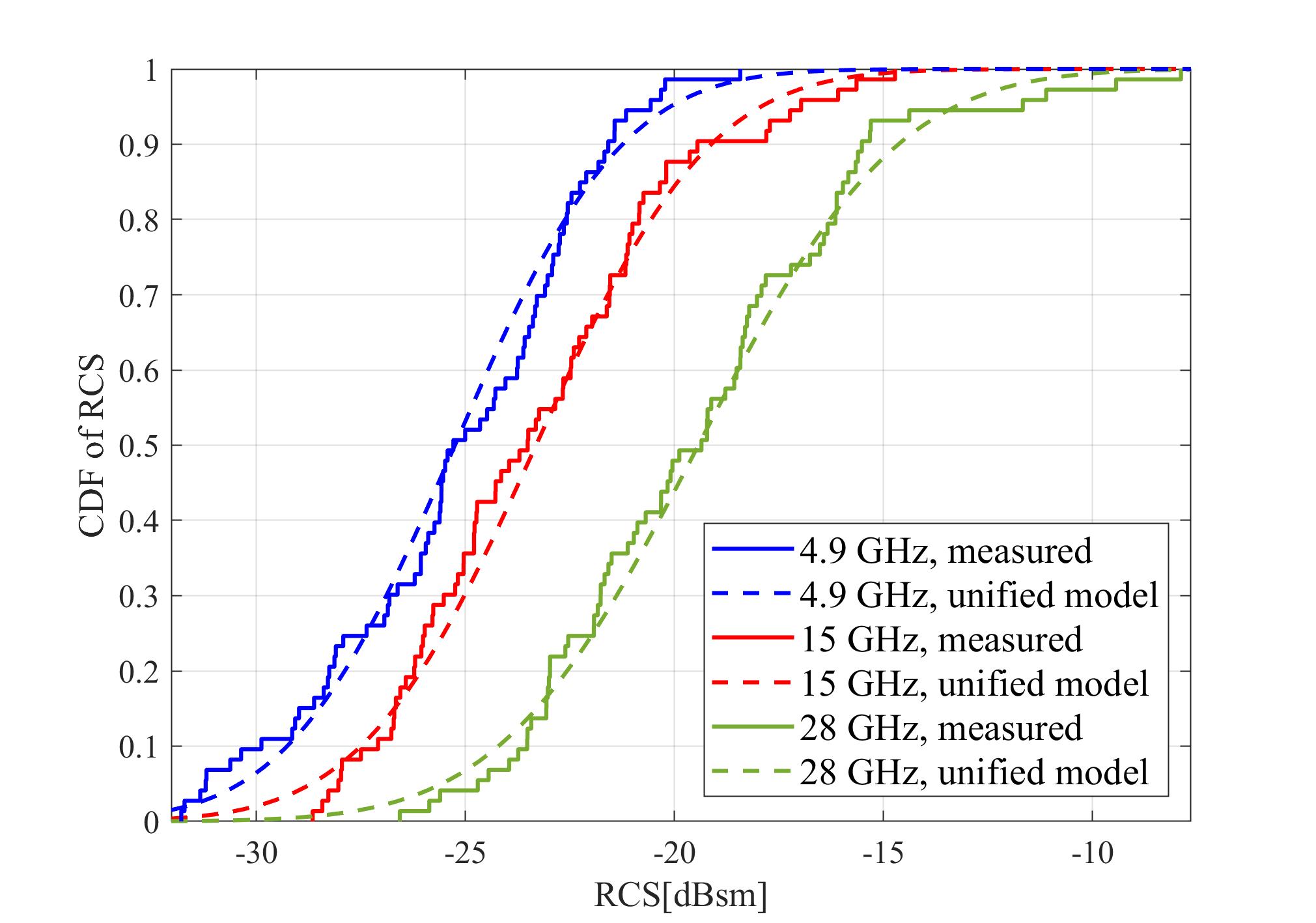}
    \caption{CDF of the UAV RCS at different frequencies.}
    \label{fig:ISAC_simulation}
\end{figure}

\begin{figure}
    \centering
    \includegraphics[width=0.95\linewidth]{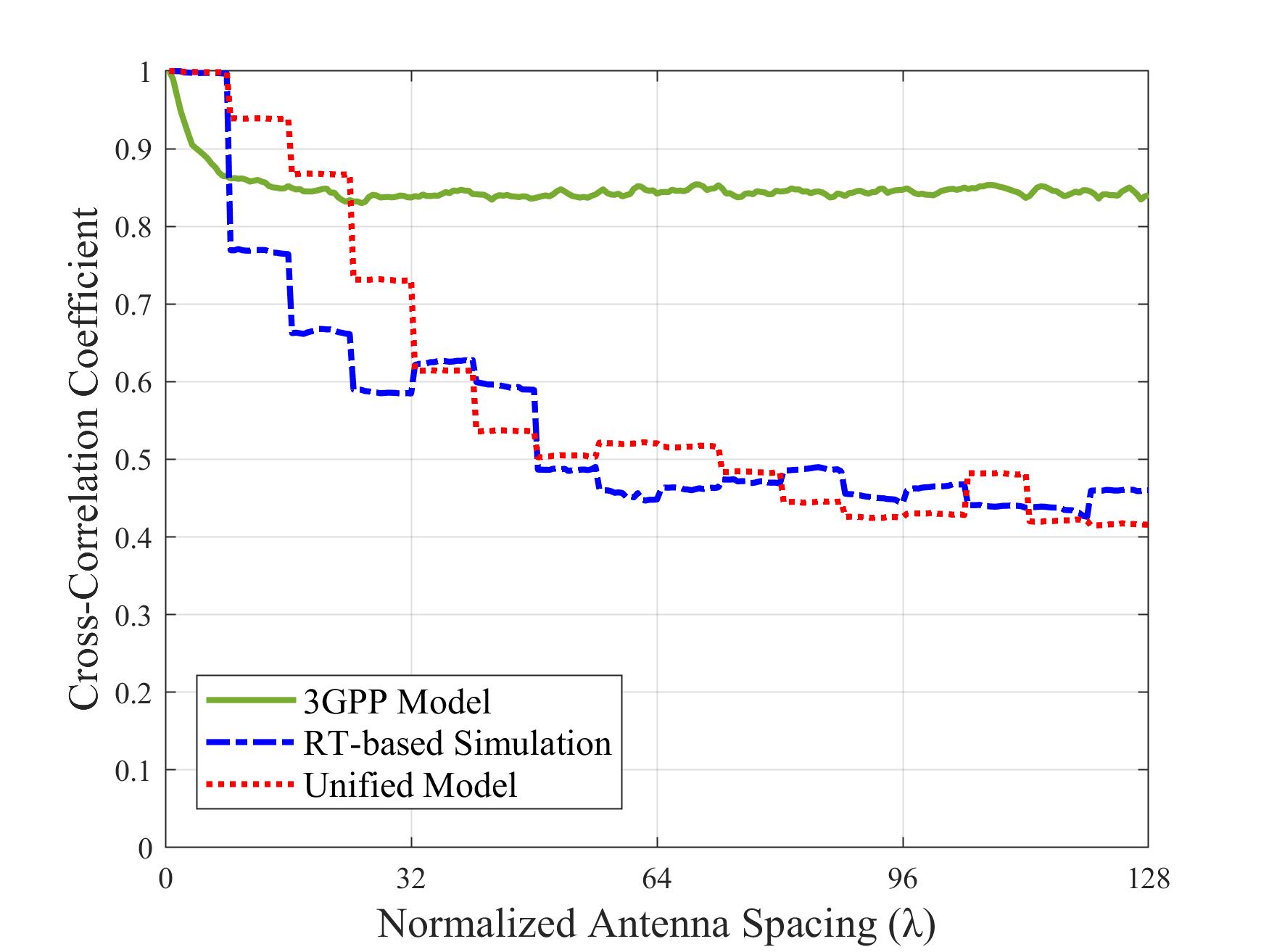}
    \caption{Cross-Correlation coefficient versus the distance between XL-MIMO antenna elements.}
    \label{fig:XLMIMO_simulation}
\end{figure}

\begin{figure}
    \centering
    \includegraphics[width=0.95\linewidth]{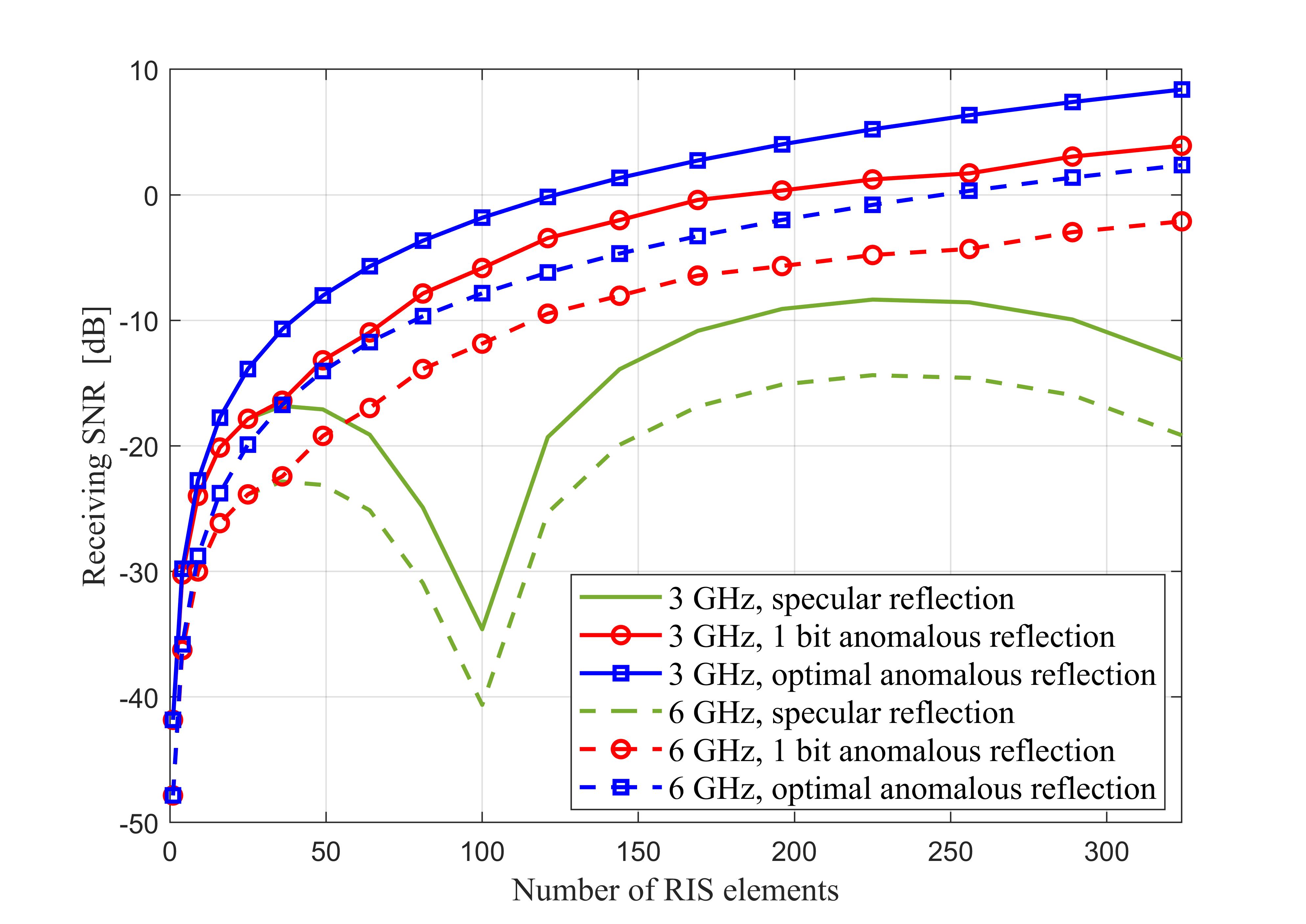}
    \caption{Receiving SNR in Tx-RIS-Rx concatenate channel with different RIS configure under unified channel model \cite{RIS_gong}.}
    \label{fig:RIS_simulation}
\end{figure}

\begin{figure}
    \centering
    \includegraphics[width=1\linewidth]{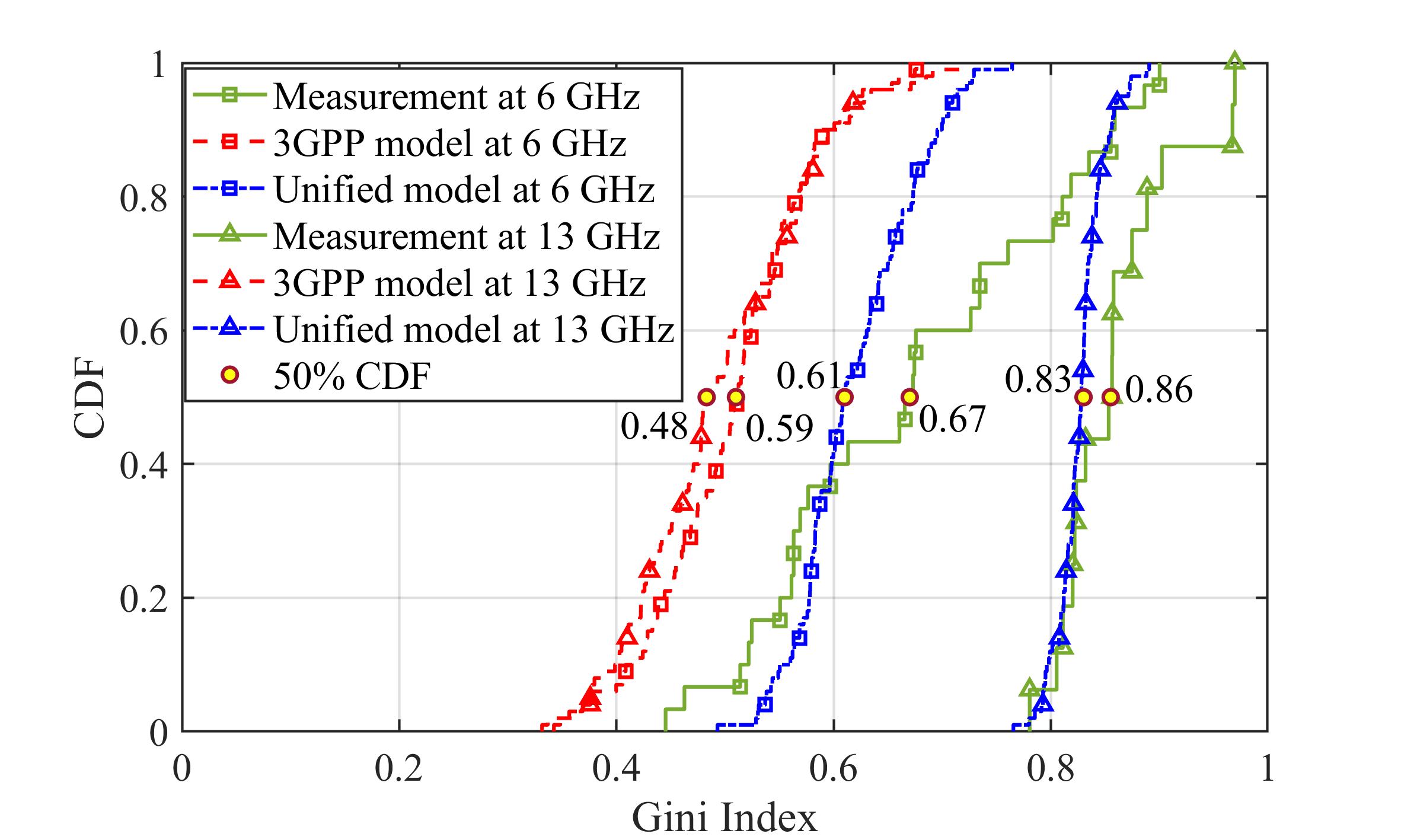}
    \caption{\textcolor{black}{Comparison of Gini coefficients between 3GPP model and unified model in new frequency band.}}
    \label{fig:Sparsity_simulation}
\end{figure}

Fig. \ref{fig:ISAC_simulation} illustrates the RCS modeling characteristics of the sensing targets in the simulation platform. The simulation includes a UAV with dimensions of $347 \times 283 \times 107$ mm, which is observed in a monostatic sensing setup. \textcolor{black}{As previously discussed, the RCS of the small UAV can be assumed isotropic in the horizontal plane and is modeled using a Gaussian distribution $N(\mu, \sigma^2)$.} The values of $\mu$ and $\sigma$ are referenced from the report of the TSG RAN\#118 meeting \cite{ISAC_3GPP_R1}. 

Fig. \ref{fig:XLMIMO_simulation} demonstrates the SnS of the XL-MIMO channel. The channel correlation coefficients of the XL-MIMO array under the original 3GPP framework and the proposed unified channel model framework are presented, with the SR length set to 8 elements in the latter. The simulation scenario is based on an InH environment. To establish a benchmark for validation, a typical InH communication scenario was constructed using ray-tracing (RT) simulation software. The results show that, unlike the traditional 3GPP model, the XL-MIMO channel correlation under the unified channel model framework decreases rapidly with increasing array distance, aligning closely with the RT simulation results, owing to the introduction of SnS.

\textcolor{black}{Fig. \ref{fig:RIS_simulation} illustrates the impact of RIS configuration on the received SNR of the concatenated Tx-RIS-Rx channel within the unified channel model framework. The positions of the transmitter, receiver, and RIS are fixed, while the RIS codebook configuration is varied, including specular reflection (codebook set to all zeros), 1-bit phase shift, and continuous phase shift configurations. 
The simulation results show that for 1-bit and continuous phase shift configurations, the received SNR increases with the number of RIS elements, and continuous phase shift achieves better performance than 1-bit phase shift.
For the specular reflection configuration, since no phase compensation is applied, the signals reflected by different RIS elements experience phase mismatches. As the RIS array size increases, the coherent addition of signals leads to fluctuations in the received power.
This indicates that the proposed simulation framework effectively captures the influence of RIS on channel performance.}

Fig. \ref{fig:Sparsity_simulation} presents the simulation of channel sparsity characteristics in new frequency bands, specifically showing the Gini coefficient results for the 6 GHz and 13 GHz bands. Simulations were conducted based on measured data, the existing 3GPP model, and the proposed unified channel model. As mentioned earlier, the Gini coefficient effectively reflects channel sparsity. The results indicate that the Gini coefficients obtained with the unified channel model are closer to the measured data, better capturing the sparsity characteristics. Additionally, the unified channel model shows an increase in channel sparsity with higher frequencies, aligning well with the measured results.

\section{Conclusion}
This article examined the state-of-the-art in 6G channel characteristics and modeling, emphasizing the challenges and requirements from a standardization perspective. To address the unique demands of 6G technologies—including ISAC, XL-MIMO, RIS, and new frequency communication—a unified channel model was proposed. The model incorporates key features such as RCS and shared scattering in ISAC, SnS and near-field effects in XL-MIMO, concatenated links and radiation patterns in RIS, and sparsity in new frequency channels. Furthermore, it supports multi-technology evaluations, wideband communications, and compatibility with existing frameworks, while leveraging measurement-based data. By providing a comprehensive and standardized framework, the proposed model aims to facilitate the development and evaluation of 6G systems.

%\section*{Acknowledgments}
%This should be a simple paragraph before the References to thank those individuals and institutions who have supported your work on this article.

\bibliographystyle{IEEEtran}
\bibliography{reference}

\begin{comment}
    \newpage

\section{Biography Section}
If you have an EPS/PDF photo (graphicx package needed), extra braces are
 needed around the contents of the optional argument to biography to prevent
 the LaTeX parser from getting confused when it sees the complicated
 $\backslash${\tt{includegraphics}} command within an optional argument. (You can create
 your own custom macro containing the $\backslash${\tt{includegraphics}} command to make things
 simpler here.)
 
\vspace{11pt}

\bf{If you include a photo:}\vspace{-33pt}
\begin{IEEEbiography}[{\includegraphics[width=1in,height=1.25in,clip,keepaspectratio]{figure/author1.jpg}}]{Huiwen Gong}
Use $\backslash${\tt{begin\{IEEEbiography\}}} and then for the 1st argument use $\backslash${\tt{includegraphics}} to declare and link the author photo.
Use the author name as the 3rd argument followed by the biography text.
\end{IEEEbiography}

\vspace{11pt}

\bf{If you will not include a photo:}\vspace{-33pt}
\begin{IEEEbiographynophoto}{John Doe}
Use $\backslash${\tt{begin\{IEEEbiographynophoto\}}} and the author name as the argument followed by the biography text.
\end{IEEEbiographynophoto}
\end{comment}

\vfill

\end{document}

%% file: table1.tex
\begin{table*}[]
\centering
\caption{\textcolor{black}{Development history of channel models from 1G to 6G.}}
\label{table1}
\resizebox{\textwidth}{!}{%
\begin{tabular}{@{}c
>{\columncolor[HTML]{F3F4F7}}c 
>{\columncolor[HTML]{eceef3}}c 
>{\columncolor[HTML]{e6e8ee}}c 
>{\columncolor[HTML]{dfe2ea}}c 
>{\columncolor[HTML]{d9dce5}}c 
>{\columncolor[HTML]{d2d6e1}}c @{}}
\toprule
\textbf{} &
  \cellcolor[HTML]{FFFFFF}\textbf{1G} &
  \cellcolor[HTML]{FFFFFF}\textbf{2G} &
  \cellcolor[HTML]{FFFFFF}\textbf{3G} &
  \cellcolor[HTML]{FFFFFF}\textbf{4G} &
  \cellcolor[HTML]{FFFFFF}\textbf{5G} &
  \cellcolor[HTML]{FFFFFF}\textbf{6G} \\ \midrule
\cellcolor[HTML]{FFFFFF}\textbf{Bandwidth}     & 30 KHz  & 200 KHz     & 5 MHz       & 100 MHz     & $\le $2 GHz   & $\le $10 GHz         \\
\cellcolor[HTML]{FFFFFF}\textbf{Frequency}     & 900 MHz & $\le $2 GHz & $\le $2 GHz & $\le $6 GHz & $\le $100 GHz & $\le $1000 GHz       \\
\cellcolor[HTML]{FFFFFF}\textbf{Enabling Tech} & FDMA    & TDMA        & CDMA        & OFDM+MIMO   & OFDM+3D MIMO  & OFDM+ISAC+XL-MIMO... \\
\cellcolor[HTML]{FFFFFF}\textbf{Dimension} &
  Time  &
  Time-frequency  &
  Time-frequency  &
  \begin{tabular}[c]{@{}c@{}}Time-frequency-Horizontal \\ angle \end{tabular} &
  \begin{tabular}[c]{@{}c@{}}Time-frequency-Horizontal\\ -Vertical angle \end{tabular} &
  \begin{tabular}[c]{@{}c@{}}Time-frequency-Horizontal\\ -Vertical angle ...\end{tabular} \\
\cellcolor[HTML]{FFFFFF}\textbf{Standard} &
   &
  COST 231 &
  \begin{tabular}[c]{@{}c@{}} 3GPP TR 25.996\end{tabular} &
  \begin{tabular}[c]{@{}c@{}}ITU-R M.2135/2155 \\ 3GPP TR 36.873\end{tabular} &
  \begin{tabular}[c]{@{}c@{}}ITU-R M.2412\\ 3GPP TR 38.900/901\\ COST 2100\end{tabular} & In developing
   \\
\cellcolor[HTML]{FFFFFF}\textbf{Model}     & Okumura-Hata  & Cost-231     & SCM       & SCME/WINNER     & GBSM/3D-GBSM   & In developing         \\
\cellcolor[HTML]{FFFFFF}\textbf{Timeline}      & 1980s   & 1990s       & 2000s       & 2010s       & 2020s         & 2030s                \\ \bottomrule
\end{tabular}%
}
\end{table*}

%% file: bare_jrnl_new_sample4.bbl
\begin{thebibliography}{10}
\providecommand{\url}[1]{#1}
\csname url@samestyle\endcsname
\providecommand{\newblock}{\relax}
\providecommand{\bibinfo}[2]{#2}
\providecommand{\BIBentrySTDinterwordspacing}{\spaceskip=0pt\relax}
\providecommand{\BIBentryALTinterwordstretchfactor}{4}
\providecommand{\BIBentryALTinterwordspacing}{\spaceskip=\fontdimen2\font plus
\BIBentryALTinterwordstretchfactor\fontdimen3\font minus \fontdimen4\font\relax}
\providecommand{\BIBforeignlanguage}[2]{{%
\expandafter\ifx\csname l@#1\endcsname\relax
\typeout{** WARNING: IEEEtran.bst: No hyphenation pattern has been}%
\typeout{** loaded for the language `#1'. Using the pattern for}%
\typeout{** the default language instead.}%
\else
\language=\csname l@#1\endcsname
\fi
#2}}
\providecommand{\BIBdecl}{\relax}
\BIBdecl

\bibitem{society_6G}
C.~D. Alwis, A.~Kalla, Q.-V. Pham, P.~Kumar, K.~Dev, W.-J. Hwang, and M.~Liyanage, ``{Survey on 6G Frontiers: Trends, Applications, Requirements, Technologies and Future Research},'' \emph{IEEE Open Journal of the Communications Society}, vol.~2, pp. 836--886, 2021.

\bibitem{ITU-R_M2160}
ITU-R, ``Framework and overall objectives of the future development of imt for 2030 and beyond,'' International Telecommunication Union, Tech. Rep. Recommendation ITU-R M.2160-0, Nov. 2023, available: https://www.itu.int/rec/R-REC-M.2160-0-202311-I.

\bibitem{38901}
3GPP, ``Study on channel model for frequencies from 0.5 to 100 ghz,'' 3rd Generation Partnership Project (3GPP), Technical Report TR 38.901 V17.0.0, Mar. 2022, available: www.3gpp.org.

\bibitem{6G_wide_band_shafi}
H.~Tataria, M.~Shafi, A.~F. Molisch, M.~Dohler, H.~Sjöland, and F.~Tufvesson, ``{6G Wireless Systems: Vision, Requirements, Challenges, Insights, and Opportunities},'' \emph{Proceedings of the IEEE}, vol. 109, no.~7, pp. 1166--1199, 2021.

\bibitem{ISAC_mag_wang}
J.~Zhang, J.~Wang, Y.~Zhang, Y.~Liu, Z.~Chai, G.~Liu, and T.~Jiang, ``{Integrated Sensing and Communication Channel: Measurements, Characteristics, and Modeling},'' \emph{IEEE Communications Magazine}, vol.~62, no.~6, pp. 98--104, 2024.

\bibitem{XLMIMO_yuan}
Z.~Yuan, J.~Zhang, Y.~Ji, G.~F. Pedersen, and W.~Fan, ``{Spatial Non-Stationary Near-Field Channel Modeling and Validation for Massive MIMO Systems},'' \emph{IEEE Transactions on Antennas and Propagation}, vol.~71, no.~1, pp. 921--933, 2023.

\bibitem{Sparsity_miao}
H.~Miao, J.~Zhang, P.~Tang, L.~Tian, X.~Zhao, B.~Guo, and G.~Liu, ``{Sub-6 GHz to mmWave for 5G-Advanced and Beyond: Channel Measurements, Characteristics and Impact on System Performance},'' \emph{IEEE Journal on Selected Areas in Communications}, vol.~41, no.~6, pp. 1945--1960, 2023.

\bibitem{Sparsity_ximan}
X.~Liu, J.~Zhang, P.~Tang, L.~Tian, H.~Tataria, S.~Sun, and M.~Shafi, ``{Channel Sparsity Variation and Model-Based Analysis on 6, 26, and 105 GHz Measurements},'' \emph{IEEE Transactions on Vehicular Technology}, pp. 1--10, 2024.

\bibitem{7-24_3GPP_M}
{3GPP TSG RAN WG1}, ``{RAN1 Chair's Notes},'' Meeting \#116-bis, Changsha, China, Apr. 2024, meeting held Apr. 15--19, 2024.

\bibitem{ISAC_liufan}
F.~Liu, Y.~Cui, C.~Masouros, J.~Xu, T.~X. Han, Y.~C. Eldar, and S.~Buzzi, ``{Integrated Sensing and Communications: Toward Dual-Functional Wireless Networks for 6G and Beyond},'' \emph{IEEE Journal on Selected Areas in Communications}, vol.~40, no.~6, pp. 1728--1767, 2022.

\bibitem{THz_guan}
B.~Peng, K.~Guan, A.~Kuter, S.~Rey, M.~Patzold, and T.~Kuerner, ``{Channel Modeling and System Concepts for Future Terahertz Communications: Getting Ready for Advances Beyond 5G},'' \emph{IEEE Vehicular Technology Magazine}, vol.~15, no.~2, pp. 136--143, 2020.

\bibitem{mid_band_rappaport}
D.~Shakya, M.~Ying, T.~S. Rappaport, H.~Poddar, P.~Ma, Y.~Wang, and I.~Al-Wazani, ``{Comprehensive FR1(C) and FR3 Lower and Upper Mid-Band Propagation and Material Penetration Loss Measurements and Channel Models in Indoor Environment for 5G and 6G},'' \emph{IEEE Open Journal of the Communications Society}, vol.~5, pp. 5192--5218, 2024.

\bibitem{platform}
{BUPT} and {CMCC}, ``{BUPT-CMCC-CMG IMT-2030 Channel Model Platform},'' [Online]. Available: \url{https://scc.bupt.edu.cn/dataset-public/datasets/22}, Feb. 2023.

\bibitem{RIS_gong}
H.~Gong, J.~Zhang, Y.~Zhang, Z.~Zhou, and G.~Liu, ``{How to Extend 3-D GBSM to RIS Cascade Channel With Non-Ideal Phase Modulation?}'' \emph{IEEE Wireless Communications Letters}, vol.~13, no.~2, pp. 555--559, 2024.

\bibitem{ISAC_3GPP_R1}
{3GPP TSG RAN WG1}, ``{ISAC Channel Measurements and Modeling},'' Meeting \#118, Maastricht, Netherlands, Aug. 2024, contribution from BUPT, CMCC, and VIVO. Meeting held Aug. 19--23, 2024.

\end{thebibliography}
